\documentclass[conference]{IEEEtran}
\usepackage{cite}

\ifCLASSINFOpdf
  \usepackage[pdftex]{graphicx}
  \graphicspath{{./eps/}{./pdf/}{./jpeg/}}
\else
\fi

\usepackage[cmex10]{amsmath}
\usepackage{cite}

\usepackage[inline]{enumitem}

\usepackage{color}
\definecolor{darkgreen}{rgb}{0, 0.75, 0}

\usepackage[noend]{algpseudocode}
\usepackage[boxed]{algorithm}

\usepackage{graphicx}
\usepackage{caption}
\usepackage{subcaption}
\usepackage{epstopdf}

\usepackage{mathtools}
\usepackage{amssymb}

\usepackage[margin=0.6745in]{geometry}

\usepackage{pgfplots}
\usepackage{xcolor}
\usepgfplotslibrary{colorbrewer}
\pgfplotsset{cycle list/Dark2}
\usetikzlibrary{matrix}
\pgfplotsset{compat=newest}
\usepgfplotslibrary{groupplots}

\usepackage{multirow}
\renewcommand{\arraystretch}{1.4}

\usepackage{float}

\usepackage{tikz}
\newcommand\copyrighttext{%
	\footnotesize \textcopyright~\the\year~IEEE. Personal use of this material is permitted.
	Permission from IEEE must be obtained for all other uses, in any current or future
	media, including reprinting/republishing this material for advertising or promotional
	purposes, creating new collective works, for resale or redistribution to servers or
	lists, or reuse of any copyrighted component of this work in other works.}

\begin{document}
\title{Routing Metrics Depending on Previous Edges:\\ The M$n$ Taxonomy and its Corresponding Solutions\vspace{-0.3cm}}
\author{\IEEEauthorblockN{Amaury Van Bemten, Jochen W. Guck, Carmen Mas Machuca and Wolfgang Kellerer}
	\IEEEauthorblockA{Lehrstuhl f\"ur Kommunikationsnetze\\Technical University of Munich, Germany\\
		Email: \{amaury.van-bemten, guck, cmas, wolfgang.kellerer\}@tum.de}}
\maketitle

\begin{abstract}
The routing algorithms used by current operators aim at coping with the demanded QoS requirements while optimizing the use of their network resources.
These algorithms rely on the \emph{optimal substructure property} (OSP), which states that an optimal path contains other optimal paths within it.
However, we show that QoS metrics such as queuing delay and buffer consumption do not satisfy this property, which implies that the used algorithms lose their optimality and/or completeness.
This negatively impacts the operator economy by causing a waste of network resources and/or violating Service Level Agreements (SLAs).
In this paper, we propose a new so-called \emph{M$n$ taxonomy} defining new metric classes.
An M$n$ metric corresponds to a metric which requires the knowledge of the $n$ previously traversed edges to compute its value at a given edge.
Based on this taxonomy, we present three solutions for solving routing problems with the newly defined classes of metrics.
We show that state-of-the-art algorithms based on the OSP indeed lose their original optimality and/or completeness properties while our proposed solutions do not, at the price of an increased computation time.
\end{abstract}


\IEEEpeerreviewmaketitle

\vspace{-0.1cm}
\section{Introduction}

		The new era of telecommunications considers new aspects such as virtual operators, large number of terminals and users and wide range of services with heterogeneous Quality of Service (QoS) requirements.
		These aspects force operators to deploy flexible networks able to guarantee all these requirements with an efficient use of their network resources.
		This paper shows that, with a strict modeling of classical QoS parameters such as delay and buffer consumption, link properties are not static and state-of-the-art algorithms lose their optimality and/or completeness properties\footnote{These properties are defined in Sec.~\ref{sec:opt-comp}.}.
    As a result, to prevent Service Level Agreements (SLAs) violations and ensure optimal use of their network resources,  operators need new solutions.

    One important challenge to ensure efficient resources usage is the optimal routing of data packets.
		Depending on the type of route that has to be found, different types of routing problems have been defined~\cite{ahuja1993network, gross2005graph}.
		For example, for a given graph, finding a path between two nodes and finding a tree from a node to a set of other nodes are two different routing problems.
    Among the set of possible solutions, it is usually also desirable to define a way for preferentially selecting one solution over another and for only selecting solutions satisfying a given set of requirements.
    For example, among the set of paths between two nodes, one might want to find the path with the greatest available bandwidth or simply to find any path with an available bandwidth greater than a given threshold.
    With this aim, so-called \emph{metric values} are associated to each edge of the graph.
    For example, if propagation delay is defined as a metric, one might want to find the path between two nodes whose total propagation delay is minimal.
    Algorithms for solving routing problems using metrics both as constraints and/or as optimization objectives have been deeply studied~\cite{paul2002survey}.

	\subsection{Motivation: Violation of the Optimal Substructure Property}\label{sec:init-motiv}

		\begin{figure}
			\centering
			\includegraphics[width=0.95\linewidth]{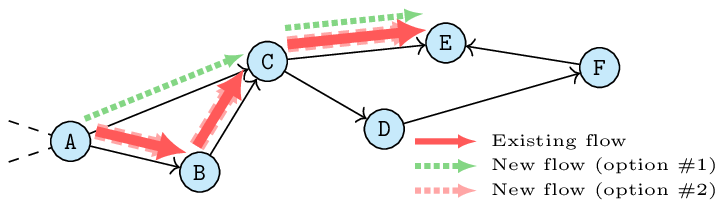}
			\vspace{-0.4cm}
			\begin{center}\large $\Downarrow$\end{center}
			\vspace{-0.1cm}
			\includegraphics[width=0.85\linewidth]{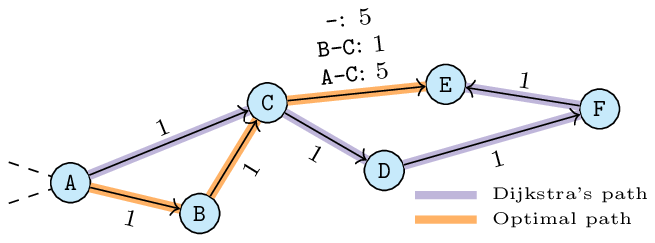}
			\caption{Example scenario involving an M$1$ metric.
      As elaborated in Sec.~\ref{sec:init-motiv}, we observe that, in such a situation, Dijkstra finds a sub-optimal path.
			}\label{queuing-example}
		\end{figure}

		State-of-the-art routing algorithms rely on the fact that the metric values associated to the edges of the graph are \emph{static} (i.e., constant) for a given routing request~\cite{ahuja1993network, gross2005graph}.
		However, this is not always the case.

		Let us examine the example depicted in Fig.~\ref{queuing-example},
		which considers as metric the queuing delay experienced at the ingress of each link (referred to as delay in this section).
		Let us assume that the network carries a flow on path \texttt{A-B-C-E} (continuous red arrows in Fig.~\ref{queuing-example}).
		Because there is only one flow in the network, no queuing occurs at any link and all the links have the same delay metric value (e.g., $1$ in this example).
		A new flow has to be routed from \texttt{A} to \texttt{E}.
		In order to reach \texttt{C}, the flow can either be routed through \texttt{A-C} or \texttt{A-B-C}.
		If it is routed through \texttt{A-C}, the flow may generate queuing at the egress of \texttt{C}, i.e., at link \texttt{C-E}, because of potential collisions of packets of the two flows coming from the two different ingress links of \texttt{C}.
		This implies that the delay metric value of \texttt{C-E} should be greater than without collisions (e.g., $5$ in this example).
		On the other hand, if the new flow is routed through \texttt{A-B-C}, it will follow the same path as the original flow, thereby generating no queuing delay at any link\footnote{We assume that the flows arrived in \texttt{A} via the same link and with the same total delay, thereby generating no queuing delay at the ingress of \texttt{A-B}.}.
		Hence, all links keep the same original delay metric value ($1$ in our case).
		We observe that the delay metric value associated to \texttt{C-E} depends on the previous edge traversed by the new flow.
		This is shown in the lower diagram of Fig.~\ref{queuing-example}.

		In such a setting, the least delay path from \texttt{A} to \texttt{E} is \texttt{A-B-C-E}, with a delay of $3$.
		However, the Dijkstra algorithm~\cite{dijkstra1959note}, an optimal algorithm for the least-delay (i.e., shortest path) problem, finds the path \texttt{A-C-D-F-E}, with a delay of $4$.
		Indeed, Dijkstra performs a breadth-first search and keeps only track of the best path to reach each node.
		While finding a path from \texttt{A} to \texttt{E}, Dijkstra will store \texttt{A-C} as the best path to reach \texttt{C} because it has a total delay of $1$, which is lower than $2$, the delay of \texttt{A-B-C}, the only other path to reach \texttt{C}.
		Then, Dijkstra will have two possibilities to reach the destination.
		Either using \texttt{C-E} with a delay of $5$, or following \texttt{C-D-F-E} with a delay of $3$.
		Since \texttt{C-D-F-E} has a lower delay, Dijkstra will choose it and return \texttt{A-C-D-F-E}, with a total delay of $4$, as final solution.
		We observe that, in this situation, Dijkstra is not able to find the optimal path.
		We will show in Sec.~\ref{loose-opt} that this is due to the fact that the \emph{optimal substructure property} (OSP), stating that an optimal path contains other optimal paths within it, is not satisfied.

    Sec.~\ref{sec:buffer-motivation} and Sec.~\ref{sec:sfc-motivation} present other scenarios for which state-of-the-art algorithms based on the OSP (i.e., nearly all state-of-the-art routing algorithms) lose either their optimality or their completeness property, or both.
    Currently, operators solve this problem by allowing sub-optimality or by having a looser modeling of QoS parameters, thereby wasting network resources, or, in the worst case, by having a too optimistic modeling of QoS parameters, thereby potentially leading to SLAs violations.
    As a result, new optimal and complete algorithms for dealing with this type of metrics are needed.

	\subsection{Contributions: M$n$ Taxonomy and Corresponding Solutions}

		The contribution of this paper is twofold.

		First, in Sec.~\ref{sec:taxonomy}, we propose the \emph{M$n$ taxonomy}, a new taxonomy defining new classes of routing metrics depending on the amount $n$ of previous edges needed to compute their value at a given edge.
		We will see that this taxonomy allows to determine whether or not state-of-the-art algorithms based on the OSP lose their optimality and/or completeness properties (see Tab.~\ref{summary-table}).

		Second, for the new classes of metrics introduced by the taxonomy, new optimal and complete algorithms are needed.
		Hence, in Sec.~\ref{sec:solutions}, we present three solutions: \emph{A*Prune}, \emph{edge-based Dijkstra} (EBD) and a \emph{graph transformation algorithm} (GTA).
		On the one hand, A*Prune~\cite{liu2001prune}, a state-of-the-art algorithm, and EBD, a newly proposed extension of the Dijkstra algorithm~\cite{dijkstra1959note}, are algorithms for the specific, respectively, \emph{multi-constrained shortest path} (MCSP) and \emph{shortest path} (SP) problems, which keep their optimality and completeness properties for the new defined classes of metrics.
		On the other hand, GTA can be used for any routing problem.
		Indeed, GTA is an extension that can be applied to any state-of-the-art algorithm for recovering its optimality and completeness properties.

		In Sec.~\ref{evaluation}, through evaluations, we show that state-of-the-art algorithms based on the OSP indeed lose their optimality and/or completeness properties depending on the M$n$ classification of the considered metric.
		Further, we show that our proposed solutions are indeed complete and optimal for their respective problems.
		However, this comes at the price of an increased runtime.

\section{Routing Metrics: Definition}\label{routing:terminology}

	Depending on the type of route that has to be found (e.g., a simple path from a single source to a single destination, multiple paths from a single source to a single destination, or a tree from a single source to several destinations), there exist a wide range of different routing problems.
	There is usually more than one possible solution to a given routing problem.
	In order to prefer one solution over another, or to provide additional requirements regarding the solutions to be accepted, so-called \emph{routing metrics} are introduced.
	Each metric defines a value, referred to as a \emph{metric value}, for each edge of the subject graph.
	These values can be updated for each routing request.
	The metrics can then be used in three different ways for selecting one or several of the available solutions.

	\subsubsection{Local Constraint Metrics}\label{sec:local}

		First, the edges that can be used by the solutions can be restricted to those satisfying a given condition based on a metric value.
		We refer to such metrics as \emph{local constraint metrics}.
		For example, in order to ensure enough bandwidth is available for a given video stream, one might want to use only links whose bandwidth is greater than the bit rate of the video stream.

	\subsubsection{Global Constraint Metrics}\label{sec:global-constraint}

		Secondly, in order to further restrict the set of solutions that can be returned, the values of a metric at each edge of a solution can be combined using a so-called \emph{link combination operator}~\cite{baumann2007survey}, e.g., the sum or the multiplication, whose result must satisfy a given constraint.
		We refer to such metrics as \emph{global constraint metrics}.
		For example, in order to ensure that the packets of a critical unicast flow arrive on time, one might want to find a path for which the sum of the delays of each of its constituting links is lower than a given threshold.

	\subsubsection{Global Optimization Metrics}\label{sec:global-optimization}

		Finally, in order to rank solutions and search for the preferred one(s), the values of a metric at each edge of a solution can be combined using a given link combination operator whose result is used for ordering the solutions.
		We refer to such metrics as \emph{global optimization metrics}.
		For example, in order to ensure that data from a unicast request is transferred as fast as possible, one might want to find the path for which the sum of the delays of each of its constituting links is minimal.

	\subsubsection{Optimality and Completeness Properties}\label{sec:opt-comp}

	An algorithm is said \emph{complete} if it always finds a solution, if one exists, satisfying both the local and global constraints.
	In case of a single global optimization metric, an algorithm is said \emph{optimal} if the solution it finds is always the optimal one.

\section{The M$n$ Taxonomy}\label{sec:taxonomy}

	In this section, we present our novel routing metric taxonomy, the \emph{M$n$ taxonomy}, classifying metrics into classes based on the amount of previous edges needed for computing their value at a given edge.
	We refer to a metric requiring the knowledge of the $n$ previously traversed edges as an M$n$ metric.

	\subsection{M$0$ Metrics: No Additional Information Required}\label{sec:m0}

		M$0$ metrics correspond to the traditional metrics considered in the state-of-the-art.
		The metric value associated to an edge depends only on the edge itself and requires no information on the other edges previously traversed.
		Examples of M$0$ metrics are the propagation delay and the total capacity of a link.

	\subsection{M$1$ Metrics: Values Depending on the Previous Edge}

		\subsubsection{Definition}

			M$1$ metrics correspond to metrics whose value at a given edge depends on the previous edge used to reach the given edge.

		\subsubsection{Motivation: Queuing Delay}

			The example developed in Sec.~\ref{sec:init-motiv} (and illustrated in Fig.~\ref{queuing-example}) corresponds to an M$1$ metric.
			Indeed, depending on which ingress link to \texttt{C} is used, the queuing delay at the egress of \texttt{C} is different.
			As a result, the metric value of \texttt{C-E} depends on the previously traversed edge and the metric is an M$1$ metric.

		\renewcommand{\arraystretch}{1.4}
		\begin{table}[t] \centering

			\begin{tabular}[width=\linewidth]{|c||c|c|c|c|}
				\hline
				/ & M$0$ & M$1$ & $\cdots$  M$n$  $\cdots$ & M$\infty$ \\\hline\hline
				\emph{Local constraint} (Sec.~\ref{sec:local}) & \textcolor{green!70!black}{\textbf{C}} \ \textcolor{green!70!black}{\textbf{O}} & \textcolor{red!90!black}{\textbf{C}} \ \textcolor{red!90!black}{\textbf{O}} & \textcolor{red!90!black}{\textbf{C}} \ \textcolor{red!90!black}{\textbf{O}} & \textcolor{red!90!black}{\textbf{C}} \ \textcolor{red!90!black}{\textbf{O}} \\\hline
				\emph{Global constraint} (Sec.~\ref{sec:global-constraint}) & \textcolor{green!70!black}{\textbf{C}} \ \textcolor{green!70!black}{\textbf{O}} & \textcolor{red!90!black}{\textbf{C}} \ \textcolor{red!90!black}{\textbf{O}} & \textcolor{red!90!black}{\textbf{C}} \ \textcolor{red!90!black}{\textbf{O}} & \textcolor{red!90!black}{\textbf{C}} \ \textcolor{red!90!black}{\textbf{O}} \\\hline
				\emph{Global optimization} (Sec.~\ref{sec:global-optimization}) & \textcolor{green!70!black}{\textbf{C}} \ \textcolor{green!70!black}{\textbf{O}} & \textcolor{green!70!black}{\textbf{C}} \ \textcolor{red!90!black}{\textbf{O}} & \textcolor{green!70!black}{\textbf{C}} \ \textcolor{red!90!black}{\textbf{O}} & \textcolor{green!70!black}{\textbf{C}} \ \textcolor{red!90!black}{\textbf{O}} \\\hline
			\end{tabular}
			\caption{Impact of M$n$ metrics on the completeness (\textbf{C}) and optimality (\textbf{O}) of state-of-the-art algorithms based on the OSP.
				\label{summary-table}}
		\end{table}

		\subsubsection{Impact as Global Optimization Metric}\label{loose-opt}

			The example developed in Sec.~\ref{sec:init-motiv} (and illustrated in Fig.~\ref{queuing-example}) corresponds to an M$1$ metric used as global optimization metric for a \emph{shortest path} (SP) problem.
			We have seen that the Dijkstra algorithm loses its optimality.
			The reason for this is that Dijkstra relies on the \emph{optimal substructure property} (OSP), which states that sub-paths of optimal paths are also optimal~\cite{thomas2009introduction}.
			While the OSP is satisfied for M$0$ metrics, it is not necessarily satisfied anymore for M$1$ metrics.
			Other routing algorithms typically also rely on the OSP, either because they are based on Dijkstra itself, or because they are based on dynamic programming or greedy approaches which are themselves based on the OSP~\cite{thomas2009introduction}.
			Consequently, other optimal SP algorithms such as Bellman-Ford~\cite{bellman1958routing, ford1956network, moore1959shortest} and A*~\cite{hart1968formal} are also affected.
			Hence, when an M$1$ metric is used as optimization metric, state-of-the-art algorithms based on the OSP lose their optimality property.
			Note that, as the metric is only used for optimization, completeness is not impacted.
			This is summarized in Tab.~\ref{summary-table} and will be confirmed in our evaluations (Sec.~\ref{sp-influence}).

		\subsubsection{Impact as Global Constraint Metric}

			Let us consider the M$1$ metric in Fig.~\ref{queuing-example} as global constraint metric (with a bound of $3.5$) and further define the hop count (an M$0$ metric) as global optimization metric.
			This corresponds to a \emph{constrained shortest path} (CSP) problem.
			CBF~\cite{widyono1994design}, an optimal CSP algorithm, is similar to Dijkstra.
			It performs a breadth-first search and keeps only track of the best path at each node.
			However, it discovers paths in order of the constraint metric and stops once the bound is reached.
			In our example, CBF would hence also find \texttt{A-C} as the best path to reach \texttt{C}.
			From this path, the destination \texttt{E} cannot be reached within the deadline.
			Hence, CBF will conclude that no path is available.
			However, \texttt{A-B-C-E}, with a delay of $3$, is a valid solution.
			As a result, CBF is incomplete.
			It can easily be shown that, with a bound of $4.5$, CBF would find \texttt{A-C-D-F-E}, with a delay of $4$, which is sub-optimal (\texttt{A-B-C-E}, with a delay of $3$, still being the optimal path).
			As for Dijkstra, this is due to the fact that the OSP is not satisfied anymore.
			Hence, other optimal CSP algorithms~\cite{guck2017unicast}, which are all based on Dijkstra and hence on the OSP, are also affected.
			Consequently, when an M$1$ metric is used as a global constraint metric, state-of-the-art algorithms based on the OSP lose both their completeness and optimality properties.
			This is shown in Tab.~\ref{summary-table} and will be confirmed in our evaluations (Sec.~\ref{csp-influence}).

		\subsubsection{Impact as Local Constraint Metric}

			It can also be easily shown that M$1$ local constraint metrics also lead to the sub-optimality and incompleteness of state-of-the-art algorithms based on the OSP (see Tab.~\ref{summary-table}).
			Indeed, the OSP is also not necessarily satisfied.

	\subsection{M$\infty$: Values Depending on the Complete Path}

		\subsubsection{Definition}
			M$\infty$ metrics correspond to metrics whose value at a given edge depends on the complete path traversed to reach the current edge.

		\subsubsection{Motivation: Buffer Management}\label{sec:buffer-motivation}

			Let us consider a metric representing the buffer consumption of a flow.
			While traversing a given path, the burstiness of a flow is increased at each hop by an amount depending on the flow and on the hop characteristics~\cite{le2001network, nc-guide}.
			The buffer consumption of a flow at a given node depends, among other things, on the burstiness of this flow.
			Hence, the buffer consumption of a flow at a given hop depends on all the previously traversed links.
			Consequently, such a metric, which should be used for per-node strict buffer management~\cite{detserv}, is an M$\infty$ metric.

		\subsubsection{Motivation: Routing through Service Function Chains}\label{sec:sfc-motivation}

			Let us consider a metric representing the total bandwidth consumption of a flow at each link.
			While routing through \emph{service function chains } (SFCs), loops can be introduced.
			Hence, the total bandwidth consumption of a flow at a given link depends on how many times the flow already visited the given link.
			As a result, all the previously visited links have to be known and this corresponds to an M$\infty$ metric.

		\subsubsection{Impact as Local Constraint Metric}\label{sec:local-constraint}

			\begin{figure}
				\centering
				\includegraphics[width=0.8\linewidth]{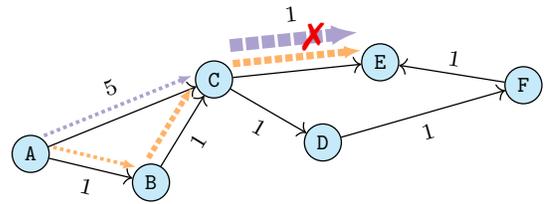}
				\caption{Example scenario involving an M$\infty$ metric.
					As elaborated in Sec.~\ref{sec:local-constraint}, we observe that the new flow will be rejected or accepted at \texttt{C-E} depending on where it is coming from.}\label{buffer-consumption-m1-metric}
			\end{figure}

			Let us consider the example developed in Sec.~\ref{sec:buffer-motivation} (illustrated in Fig.~\ref{buffer-consumption-m1-metric}) and the problem of routing a flow from \texttt{A} to \texttt{E} (Fig.~\ref{buffer-consumption-m1-metric}) through the least-hop path (M$0$ metric).
			We consider a local constraint metric rejecting flows consuming too much buffer space.
			In Fig.~\ref{buffer-consumption-m1-metric}, each link is labeled with its queuing delay (generated by already embedded flows).
			Note that this queuing delay is not used as a metric.
			As per network calculus~\cite{le2001network, nc-guide}, the burstiness of the flow, i.e., its buffer consumption, will increase approximately proportionally to the delay it experiences along its path.
			Fig.~\ref{buffer-consumption-m1-metric} shows the two options for routing the new flow.
			The width of the arrows represent the buffer consumption of the flow.
			While routing, Dijkstra will save the path \texttt{A-C}, with a hop count of $1$, as the best path towards \texttt{C}.
			However, because the queuing delay experienced at \texttt{A-C} is $5$, the burstiness of the flow greatly increased and it cannot be accepted at link \texttt{C-E}.
			As a result, Dijkstra would either find no solution (thereby losing completeness) or find \texttt{A-C-D-F-E}, with a hop count of $4$, if the latter has enough buffer space available.
			However, \texttt{A-B-C-E}, with a hop count of $3$, has a low delay, thereby only slightly increasing the burstiness of the flow and hence allowing it to use \texttt{C-E}.
			In this case, Dijkstra is hence sub-optimal.
			This is again due to the fact that the OSP is not satisfied.

			\begin{figure}
				\centering
				\vspace{-0.4cm}
				\includegraphics[width=0.6\linewidth]{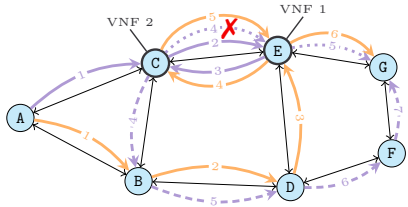}
				\caption{Example scenario involving an M$\infty$ metric.
        As elaborated in Sec.~\ref{sec:local-constraint}, the bandwidth consumption of the new flow at a link depends on whether or not it visited this link before.
        Hence, acceptance of the new flow at a link depends on all the previously visited links.}\label{bandwidth-consumption-minf-metric}
			\end{figure}
			Let us further consider the example developed in Sec.~\ref{sec:sfc-motivation} (illustrated in Fig.~\ref{bandwidth-consumption-minf-metric}) and the problem of routing a flow on the least-hop path (M$0$ metric) from \texttt{A} to \texttt{G} (Fig.~\ref{bandwidth-consumption-minf-metric}) visiting the two \emph{virtual network functions} (VNFs) \texttt{E} and \texttt{F} in the specified order.
			The flow consumes $700$ Mbps and each link has $1$ Gbps available bandwidth.
			Hence, the flow can only visit each link once.
			A least-hop algorithm would first reach the first VNF, i.e., \texttt{E}, through the least-hop path, i.e., \texttt{A-C-E}.
			Then, it would visit the second VNF, i.e., \texttt{C}, through the least-hop path from \texttt{E}, i.e., \texttt{E-C}.
			Finally, it would try to reach the destination with the least-hop path from \texttt{C}, i.e., \texttt{C-E-G}.
			However, this would imply visiting a second time \texttt{C-E}, which is not allowed.
			If the algorithm does not notice that the access to \texttt{C-E} is refused (e.g., if it cached the result of the access control when it first traversed the link), it will return an invalid path and hence be incomplete.
			If the algorithm notices that the access to \texttt{C-E} is refused, it would follow the second least-hop path, i.e., \texttt{C-B-D-F-E}, thereby finding \texttt{A-C-E-C-B-D-F-E}, with a hop count of $7$, as final solution.
			However, \texttt{A-B-D-E-C-E-G} is a valid path with a hop count of $6$.
			Hence, the algorithm loses its optimality.

			As a result, when an M$\infty$ metric is used as a local constraint metric, state-of-the-art algorithms based on the OSP lose both their completeness and optimality properties (Tab.~\ref{summary-table}).

		\subsubsection{Impact as Global Optimization or Constraint Metric}

			As an M$1$ metric is also an M$\infty$ metric, it can easily be shown that the impact as global optimization and constraint metric is the same for both M$1$ and M$\infty$ metrics (Tab.~\ref{summary-table}).

	\subsection{M$n$: Values Depending on the $n$ Previous Edges}

	As a generalization, we further introduce the class of M$n$ metrics.
	M$n$ metrics correspond to metrics whose value at a given edge depend on the $n$ previous edges used to reach the current edge.

	Obviously, for the same reasons as for M$\infty$ metrics, the impact of state-of-the-art algorithms based on the OSP is the same as for M$1$ metrics.
	This is shown in Tab.~\ref{summary-table}.

\section{Solutions for the M$n$ Taxonomy}\label{sec:solutions}

	In this section, we present three opportunities (Sec.~\ref{sec:aprune}, \ref{sec:ebd} and \ref{sec:gta}) to optimally and completely solve problems with M$n$ metrics, $n > 0$.

	\subsection{Existing Solution: A*Prune}\label{sec:aprune}

		\emph{A*Prune}~\cite{liu2001prune} is a complete and optimal state-of-the-art algorithm able to solve the \emph{shortest path} (SP) and \emph{multi-constrained shortest path} (MCSP) problems.
		Although similar to Dijkstra, A*Prune does not rely on the OSP but is only faster when it is satisfied.
		Hence, it keeps its optimality and completeness properties for both M$n$ and M$\infty$ metrics.
		The reason for this is that it does not keep track of only one best path to reach each node.
		Instead, all feasible paths are kept in memory and the best ones are extended first.
		Path extension is stopped only once the next path to extend has an optimization metric value higher than the current best path for the destination.

		In the example of Fig.~\ref{queuing-example}, as Dijkstra, A*Prune will first find the path \texttt{A-B-C-E} with a total metric value of $7$.
		However, the next path to extend, namely \texttt{A-C}, has a total metric value of $4$, which is lower than the current best path to the destination.
		Hence, A*Prune will further extend \texttt{A-C} and thereby find \texttt{A-C-E}, which is optimal.

		Unfortunately, this optimality for any type of M$n$ and M$\infty$ metric comes at the price of a poor scalability behavior.
		This will be shown in our evaluations (Sec.~\ref{evaluation}).

	\subsection{New Solution: Edge-based Dijkstra (EBD)}\label{sec:ebd}

		In the particular case of the SP problem with an M$1$ optimization metric, Dijkstra can be slightly adapted.
		Instead of keeping track of the best path towards each node, our proposed adaptation keeps track of the best path towards each \emph{edge}.
		We refer to this algorithm as \emph{edge-based Dijkstra} (EBD).

		In the example of Fig.~\ref{queuing-example}, instead of keeping track of the best path towards node \texttt{C}, EBD will keep track of the best path towards \texttt{A-C} (which is \texttt{A-C} itself) and towards \texttt{B-C} (which is \texttt{A-B-C}).
		Then, when extending these paths to obtain the best path towards \texttt{C-E}, both paths will be considered and the path \texttt{A-C-E}, missed by the normal Dijkstra, will be found.
		Once EBD stops, the final solution then corresponds to the best path among those stored at all the ingress links to the destination node.

		Because the amount of edges in a graph is usually higher than the amount of nodes, EBD keeps track and extends more paths than the traditional Dijkstra and the usage of EBD for M$1$ metrics hence results in a runtime increase compared to a normal Dijkstra run.
    This will be confirmed in Sec.~\ref{sp-influence}.

	\subsection{New Solution: Graph Transformation Algorithm (GTA)}\label{sec:gta}

		In this section, we propose a \emph{graph transformation algorithm} (GTA) which transforms a graph with M$n$ metrics to an equivalent graph with M$0$ metrics such that any routing problem with M$n$ metrics can be solved with a state-of-the-art algorithm for the given routing problem with traditional M$0$ metrics.

		\subsubsection{Reasoning}

			For M$1$ metrics, edges can have as many different metric values as ingress links (plus one for the ``\emph{null}'' ingress link when the flow starts at the given edge).
			This is shown in Fig.~\ref{queuing-example}.
			The idea of GTA is to duplicate links as many times as they have different metric values such that each new edge has a static M$0$ metric value.
			Let us refer to the original graph with M$n$ metrics as the \emph{M$n$ graph} and to the transformed graph with only M$0$ metrics as the \emph{M$0$ graph}.
			Each edge in the M$0$ graph then corresponds to \emph{(i)} an edge of the original M$n$ graph, and \emph{(ii)} a set of $n$ previous edges.
			In this way, edges in the M$0$ graph have only one metric value (i.e., the metric is now an M$0$ metric) and state-of-the-art algorithms based on the OSP can operate properly.

		\subsubsection{Algorithm Description for M$1$ Metrics}\label{gta-m1-sec}

			\begin{figure}
				\centering
				\vspace{-0.4cm}
				\includegraphics[width=0.7\linewidth]{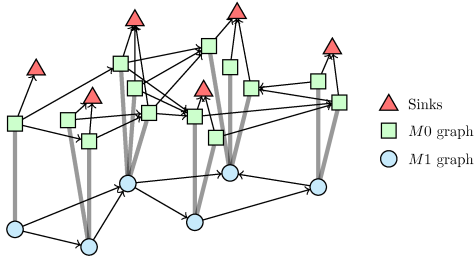}
				\caption{Illustration of the GTA procedure for M$1$ metrics.
					From the original graph with M$1$ metrics (the M$1$ graph), the algorithm creates a new graph (the M$0$ graph) with M$0$ metrics on which any state-of-the-art algorithm can run to solve the original M$1$ problem.}\label{gta}
			\end{figure}

			The GTA algorithm for M$1$ metrics is illustrated in Fig.~\ref{gta}.
			The blue circle nodes correspond to the original M$1$ graph.
			From this graph, each node is copied and then duplicated as many times as it has ingress links.
			Each M$0$ node (green square nodes in Fig.~\ref{gta}) then corresponds to an original M$1$ node and to one ingress link of this node (including the ``\emph{null}'' ingress link).
			Then, M$0$ edges are obtained by creating edges towards the created M$0$ nodes from all the M$0$ nodes corresponding to the source of the original edge to which the destination M$0$ node corresponds.
			Each M$0$ edge then corresponds to \emph{(i)} an edge of the M$1$ graph, and \emph{(ii)} an ingress link to this edge.
			Hence, each M$0$ edge can be assigned a static M$0$ metric value corresponding to the metric value of the original edge when the given ingress edge is used to reach it.
			We refer to this procedure as \textsc{GTA()}.

		\subsubsection{Request and Result Transformation}
			In order for the original algorithm to run on the transformed graph, the original request has to be mapped.
			First, the source nodes now correspond to their M$0$ equivalent which have no ingress link.
			Secondly, the destination nodes have now several M$0$ equivalents.
			To overcome this problem, so-called \emph{sink nodes} have to be created.
			All the M$0$ nodes corresponding to the same original M$1$ node have to be connected to the same sink node with edges whose metric value does not change the metric value of the overall solution (that is, e.g., $0$ for an additive metric).
			The destination(s) of the original request then become(s) the corresponding sink node(s).
			We refer to the procedure of creating the sink nodes as \textsc{AddSinks()}.
			Once the algorithm found a solution on the M$0$ graph, the solution on the M$1$ graph can be recovered by taking all the M$1$ equivalents of the elements of the M$0$ solution returned by the algorithm.

		\subsubsection{Algorithm Description for M$n$ Metrics}

			\begin{figure}
				\centering
				\includegraphics[width=0.45\linewidth]{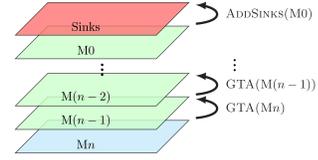}
				\caption{Illustration of the GTA procedure for M$n$ metrics.
					The procedure described in Fig.~\ref{gta} simply has to be applied $n$ times, provided that the sinks are only added at the end.}\label{gta-n}
			\end{figure}

			In order to transform a graph with M$n$ metrics, the \textsc{GTA()} procedure described in Sec.~\ref{gta-m1-sec} simply has to be applied $n$ times and followed by the addition of sink nodes (Fig.~\ref{gta-n}).
			Indeed, as \textsc{GTA()} duplicates edges for each ingress link, applying it $n$ times will duplicate edges for each possible set of $n$ ingress links and hence lead to an M$0$ graph.

		\subsubsection{Cost of the Transformation}\label{sec:complexity-increase}

\begin{figure}
		\centering
		\vspace{+0.05cm}
		\includegraphics[width=0.5\linewidth]{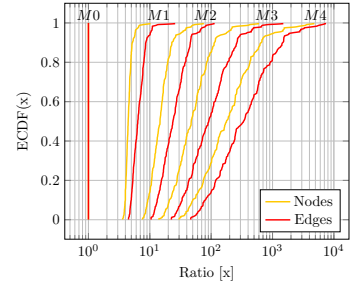}
		\vspace{-0.0cm}
		\caption{Evolution of the amount of nodes and edges for the Internet Topology Zoo~\cite{topology-zoo} topologies for different amount of executions of GTA (including the sinks creation).}\label{complexity-increase}
	\end{figure}

			The size of the M$0$ graph increases with $n$.
			Fig.~\ref{complexity-increase} shows the evolution of the amount of nodes and edges for all the topologies from the Internet Topology Zoo~\cite{topology-zoo}.
			We can observe that the amount of nodes and edges increases by up to one order of magnitude for each application of the \textsc{GTA()} procedure.
			That is, while GTA allows to optimally solve any problem with M$n$ metrics using algorithms for M$0$ metrics, this comes at the price of a huge increase in the graph size.
			An insight on the runtime impact will be given in Sec.~\ref{evaluation}.

\vspace{-0.06cm}
\section{Evaluation}\label{evaluation}

	The goal of the evaluation is to observe the impact of M$1$ and M$\infty$ metrics on the optimality and completeness of state-of-the-art algorithms based on the OSP and to show that our proposed solutions are correct.
	In particular, we show the influence of M$1$ and M$\infty$ metrics on one SP and one CSP algorithm both with and without GTA.
	Algorithms are compared to A*Prune, which provides a benchmark for both completeness and optimality.

	\subsection{Setup}

		\subsubsection{Topologies}\label{se:topo}

		We use the topologies from the Internet Topology Zoo~\cite{topology-zoo} which are connected and have more than 10 nodes.
		Further, because A*Prune poorly scales both in terms of memory consumption and runtime~\cite{guck2017unicast}, we filter out topologies with more than 100 nodes or 200 edges.

		\subsubsection{Requests}
		For a given topology, source and destination nodes for requests are randomly selected from the whole set of nodes of the topology.

		\subsubsection{Metrics}
		We define one M$0$, one M$1$ and one M$\infty$ metric.
		The metric values are random values between $1$ and $2$.
		For the M$0$ metric, the values are defined for each edge.
		For the M$1$ metric, the values are defined for each combination of edge and previous edge.
		For the M$\infty$ metric, the values are defined for each path.

	\subsection{Measurement Environment}

		In both SP and CSP scenarios, each algorithm is ran 20.000 times for each topology and metric type.
		Prior to these 20.000 runs, 1.000 warm-up runs are used to prevent the Java Hotspot optimizer from influencing the runtime measurements.
		The evaluation ran on an Intel Core i7-4790 CPU @ 3.60GHz.

	\subsection{Shortest Path: Optimality Influence}\label{sp-influence}

		\begin{figure}
			\centering
			\vspace{-0.5cm}
			\includegraphics[width=\linewidth]{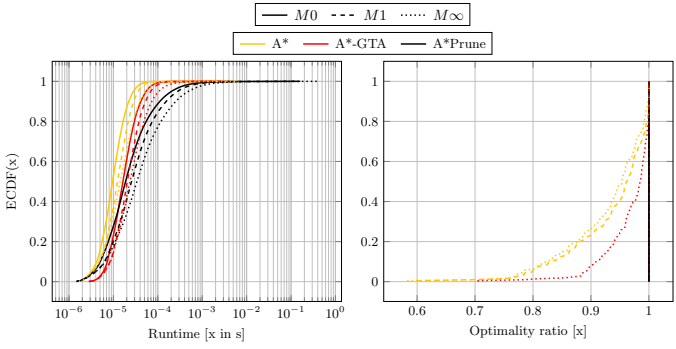}
			\vspace{-0.4cm}
			\caption{Runtime and optimality ratio of A*, A* with one GTA transformation (A*-GTA) and A*Prune for the SP problem with M$0$, M$1$ and M$\infty$ optimization metrics.}
			\vspace{-0.1cm}
			\label{optimality}
		\end{figure}

		\subsubsection{Setup}

			We observe the runtime and optimality of A*~\cite{hart1968formal} (an optimal SP algorithm) and A* with GTA applied once (referred to as A*-GTA) for M$0$, M$1$ and M$\infty$ metrics using A*Prune as a benchmark.
			The guess values for both A*Prune and A* correspond to the hop count.

		\subsubsection{Results: Runtime}

			The left plot of Fig.~\ref{optimality} shows the empirical cumulative distribution function (ECDF) of the observed running times during the simulation\footnote{Note that, for GTA, the graph creation is not taken into account. Indeed, it can be done once for all the requests.}.
			We can observe that the GTA transformation for A* leads to an increase in runtime of around half an order of magnitude.
			Even though big topologies were not used, the figure also illustrates the poor scalability of A*Prune.
			Indeed, while A*Prune is sometimes significantly faster than A*-GTA (around 40\% of the cases for the M$0$ metric, around 30\% of the cases for the M$1$ metric and around 20\% of the cases for the M$\infty$ metric) its runtime becomes very high for bigger topologies.
			Note that the runtime increase for each algorithm for the M$1$ and M$\infty$ metrics is mostly due to the increased complexity of the metric values computation.

		\subsubsection{Results: Optimality Ratio}

			The right plot of Fig.~\ref{optimality} shows the ECDF of the \emph{optimality ratio} observed for each algorithm and metric types.
			The optimality ratio is defined per topology as the percentage of requests that the algorithm was able to solve optimally.
			As expected, both A* and A*-GTA are always optimal for M$0$ metrics and show a sub-optimal behavior for M$\infty$ metrics.
			However, while A* presents a sub-optimal behavior for M$1$ metrics, A*-GTA does not.
			The sub-optimality of A* for M$1$ metrics corresponds to the behavior described in Sec.~\ref{sec:init-motiv} and Fig.~\ref{queuing-example}.
			This confirms that GTA allows state-of-the-art algorithms based on the OSP to keep their original properties for M$1$ metrics.
			For M$\infty$ metrics, though GTA does not guarantee optimality for A*, we can observe that it improves its optimality ratio.
			All algorithms were complete, confirming that an M$1$ or M$\infty$ global optimization metric does not impact the completeness of algorithms (see Tab.~\ref{summary-table}).

		\subsubsection{Results: Conclusions}

			M$1$ and M$\infty$ global optimization metrics indeed lead to the sub-optimality of A*.
			While A*Prune provides optimality for any type of metric, it presents a problematic scalability behavior.
			For its part, GTA allows A* to optimally solve problems for M$1$ metrics and improves its optimality ratio for M$\infty$ metric, at the price of a reasonable increased runtime.
			Note that EBD led to an identical optimality behavior and a similar runtime behavior as A*-GTA.

	\subsection{Constrained Shortest Path: Completeness Influence}\label{csp-influence}

		\begin{figure}
			\centering
			\vspace{-0.6cm}
			\includegraphics[width=0.97\linewidth]{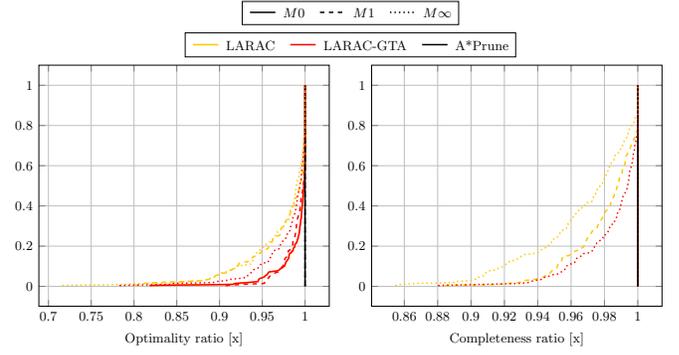}
			\vspace{+0.0cm}
			\caption{Optimality and completeness ratios of LARAC, LARAC with one GTA transformation (LARAC-GTA) and A*Prune for the CSP problem with M$0$, M$1$ and M$\infty$ global constraint metrics and an M$0$ optimization metric.
			}
			\vspace{-0.1cm}\label{completeness}
		\end{figure}

		\subsubsection{Setup}

			Because CBF, the optimal CSP algorithm, presents an exponential runtime behavior~\cite{widyono1994design}, we use LARAC~\cite{aneja1978constrained, handler1980dual, blokh1996approximate, juttner2001lagrange}, a sub-optimal but fast and complete CSP algorithm.
			We then observe the optimality and completeness of LARAC and LARAC with GTA applied once (referred to as LARAC-GTA) for M$0$, M$1$ and M$\infty$ metrics, using A*Prune as benchmark.
			Note that, because of the poor scalability of A*Prune and the higher runtime required by a CSP search compared to an SP search, we further reduced the topologies to those with less than 50 nodes and 100 edges.

		\subsubsection{Metrics and Constraint Bounds}

			For the optimization metric, we use an M$0$ metric.
			The constraint metric upper bound is randomly distributed among all the possible values (between the minimum and the maximum values).

		\subsubsection{Results: Runtime}

			Because of space constraints and because the runtime impact of GTA appeared to be the same as for the SP problem, we omit the runtime values for the CSP problem.

		\subsubsection{Results: Optimality Ratio}

			The left plot of Fig.~\ref{completeness} shows the ECDF of the optimality observed for each algorithm.
			LARAC and LARAC-GTA present exactly the same behavior for the M$0$ metric (the yellow curve being hidden in Fig.~\ref{completeness}).
			As expected, LARAC is not optimal for the M$0$ metric.
			For LARAC, the M$1$ metric reduces its optimality ratio.
			However, the M$\infty$ metric does not appear to further reduce this optimality ratio.
			For LARAC-GTA, the M$1$ metric does not have a big influence on its optimality.
			However, the M$\infty$ metric appears to further reduce this optimality ratio.
			It is interesting to notice that this behavior is different for LARAC and LARAC-GTA.

		\subsubsection{Results: Completeness Ratio}

			The right plot of Fig.~\ref{completeness} shows the ECDF of the \emph{completeness ratio} observed for each algorithm.
			The completeness ratio is defined per topology as the percentage of requests for which the algorithm was able to find a solution.
			The same conclusions as for the optimality ratio of the SP simulations can be drawn.
			Indeed, GTA allows LARAC to be complete for the M$1$ metric and improves on its completeness ratio for the M$\infty$ metric.

		\subsubsection{Results: Conclusions}

			The M$1$ and M$\infty$ global constraint metrics indeed lead to the incompleteness of LARAC.
			GTA allows LARAC to completely solve problems for the M$1$ metric and improves its completeness ratio for the M$\infty$ metric.

\section{Applicability of the Solutions}\label{other-problems}

	A*Prune and EBD can only be used for the SP/MCSP and SP problems, respectively.
	On the other hand, GTA can be applied to any routing algorithm.
	That is, GTA can be used for any routing problem (unicast, multicast, multipath, etc.) and amount of optimization and constraint metrics as long as a state-of-the-art algorithm for the corresponding problem with M$0$ metrics exists.
	However, we identify that multipath routing requires small adaptations.
	Indeed, after a GTA transformation, disjointness on the transformed graph does not guarantee disjointness of the corresponding elements on the original graph and multipath routing algorithms might hence return disjoint paths on the transformed graph which are not disjoint on the original graph.
	To circumvent this problem, the GTA transformation can be adapted by adding an intermediate \emph{sink edge} to which all the transformed edges corresponding to an identical original edge connect.
	In this way, if the algorithm finds two paths that use different edges corresponding to the same original edge, it will have to use the same sink edge and will hence conclude that these paths are not disjoint.
	This additional procedure however again comes at the price of an increased amount of nodes and edges in the transformed graph.

\section{Conclusions}

    State-of-the-art routing algorithms are optimal and complete when using metrics that satisfy the \emph{optimal substructure property} (OSP).
    However, we have shown that relevant QoS metrics such as delay or buffer consumption do not satisfy this property.
    Hence, the algorithms lose their optimality and/or completeness.
    This causes operators to violate Service Level Agreements (SLAs), hence incurring penalties, as well as to inefficiently use their network resources.

    In order to still guarantee optimal and complete results, we first proposed a new M$n$ metric taxonomy for classifying routing metrics based on the amount $n$ of previously traversed edges needed to compute their value at a given edge.
    Based on this taxonomy, we presented solutions guaranteeing optimality and completeness.
    First, we presented \emph{A*Prune}~\cite{liu2001prune}, a state-of-the-art algorithm that can deal with any type of M$n$ and M$\infty$ metric for solving the shortest path (SP) and multi-constrained shortest path (MCSP) problems.
    Second, we proposed \emph{edge-based Dijkstra} (EBD), a newly proposed modification of Dijkstra for solving SP problems with M$1$ metrics.
    Finally, because A*Prune and EBD can only be used for particular problems and metric types, we proposed a \emph{graph transformation algorithm} (GTA) that allows any state-of-the-art algorithm for any routing problem (e.g., unicast, multicast, multi-constrained, etc.) to solve problems with M$n$ metrics.
    While A*Prune is the only opportunity for optimally solving a problem with M$\infty$ metric, we have shown that it presents a poor scalability behavior.
    Besides, on the example of the A*~\cite{hart1968formal} and LARAC~\cite{aneja1978constrained, handler1980dual, blokh1996approximate, juttner2001lagrange} state-of-the-art algorithms, we have shown that GTA indeed recovers their properties for M$n$ metrics, at the cost of an increased running time.

    The effective impact of the completeness and optimality behavior of state-of-the-art algorithms and of the proposed algorithms on the network resources usage of operators is an interesting future research direction and is left for future work.

\section*{Acknowledgments}

This work has received funding from the European Union's Horizon 2020 research and innovation programme under grant agreement No.~671648 (VirtuWind).

\end{document}